# Ferroelectric Polarization in Antiferroelectric Chalcogenide Perovskite BaZrS$_3$ Thin Film

Juhi Pandey,[1] Debjit Ghoshal,[2] Dibyendu Dey,[3,4] Tushar Gupta,[5] A. Taraphder,[3] Nikhil Koratkar,[5] and Ajay Soni[1]*

[1] School of Basic Sciences, Indian Institute of Technology Mandi, Mandi-175075, Himachal Pradesh, India

[2] Department of Chemical and Biological Engineering, Rensselaer Polytechnic Institute, Troy-12180, New York, USA

[3] Department of Physics and Centre for Theoretical Studies, Indian Institute of Technology Kharagpur, Kharagpur-721302, India

[4] Department of Physics, Arizona State University, Tempe-85287, Arizona, USA

[5] Department of Mechanical, Aerospace and Nuclear Engineering, Rensselaer Polytechnic Institute, Troy-12180, New York, USA.

Corresponding Author Email: ajay@iitmandi.ac.in

**ABSTRACT:**

Bulk chalcogenide perovskite BaZrS$_3$ (BZS), with a direct band gap in visible region, is an important photovoltaic material, albeit with limited applicability owing to its antiferroelectric (AF) nature. Presently, ferroelectric (FE) perovskite-based photovoltaics are attracting enormous attention for environmental stability and better energy conversion efficiency through enhanced charge separation, owing to loss of center of inversion symmetry. We report on antiferroelectric-ferroelectric (AF-FE) phases of BZS thin film, grown with chemical vapor deposition (CVD), using temperature-dependent Raman investigations and first-principles calculations. The origin of FE phases is established from anomalous behavior of $A_g^7$ ~ 300 cm$^{-1}$ and $B_{1g}^5$ ~ 420 cm$^{-1}$ modes, which involves the vibration of atoms at apical site of ZrS$_6$ octahedra. Additionally, below 60 K, $B_{1g}^1$ and $B_{2g}^2$ ( ~ 85 cm$^{-1}$) modes appear whereas $B_{2g}^1$ (~ 60 cm$^{-1}$) disappears to stabilize the *Pnma* structure against ferroelectricity by local distortion. Here, $B_{2g}^2$ and $B_{2g}^1$ involve vibrations of Ba atoms in AF manner while $B_{1g}^1$ involves, in addition, the rotation of octahedra as well. Our first-principles calculations confirm that FE appears as a result of loss of center of inversion symmetry in ZrS$_6$ octahedra due to existence of oxygen (*O*) impurities placed locally at apical sites of sulfur (*S*) atom.

Perovskite materials have remarkable technological importance in the field of photovoltaics, optoelectronics, energy harvesting, ferroelectricity, electrocatalysis, photocatalysis, magnetic storage and spintronics because of their exotic optical, mechanical, magnetic and electrical transport properties.[1-5] Several transition metal chalcogenide perovskites (TMCPs), such as $CaZrS_3$, $BaZrS_3$, $SrZrS_3$ have emerged as non-toxic and environmentally stable photovoltaic material with electronic band gap in the visible region.[1-2, 6-7] Further, to achieve optimum photovoltaic performance, the electronic band gap can be modulated by alloying and doping. Commonly, TMCPs are distorted perovskites with either AF or paraelectric (PE) properties. Therefore, engineering ferroelectricity by designing materials is necessary to develop promising FE TMCPs. Noticeably, there are several ways to induce AF-FE phase in AF materials owing to comparable energy of AF and FE phases.[8] Thin films of AF materials below a critical thickness induces FE phase due to surface effect, epitaxial strain and dimensionality of nanostructures.[8] For instance, bulk $PbZrO_3$ is AF in nature but single crystalline films below ~ 22 nm thickness have demonstrated FE phases due to epitaxial strain.[9] Additionally, FE phases in thin films can also be induced by applications of large electric field which generally results in large lattice strain and high electrostatic energy density. Thus, the pairing of epitaxial thin film of FE and AF materials are gaining increasing consideration for integration into standard semiconductor technology.[10]

Bulk BZS is an AF chalcogenide perovskite having direct band gap in the range of 1.7-1.85 eV and has strong absorption in the visible spectrum.[2] The band gap in BZS can be tuned by cation or anion alloying, and fractional substitution of Zr atoms for possible application in single junction solar cell.[11] Being similar in crystal structure to perovskite oxides such as $BaTiO_3$, $SrTiO_3$, $KTaO_3$ and $KNbO_3$, BZS may also demonstrate AF to FE phase transition with temperature. However, bulk BZS shows long term stability against environmental conditions at low temperature and high pressure, whereas oxide have very high band gap and alternative - halide perovskites are unstable against surface oxidation and moisture.[7, 12-13] Additionally, thin films of BZS can show FE phases and modified functional properties due to strain, lattice mismatch with substrate, impurities, defects and superlattice formation. In this regards, unlike the established understanding of phase transition in bulk TMCPs, thin films of perovskites are not yet well-explored and tuning of FE-AF-PE nature by replacement of atoms at anion site for TMCPs is an open challenge.

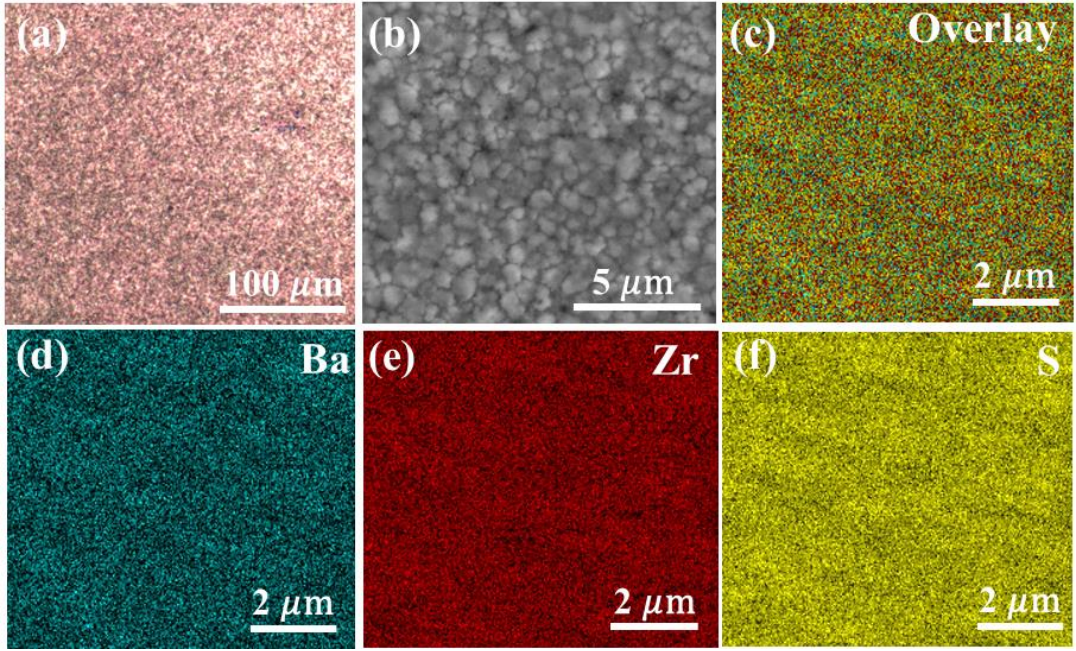

**Figure 1**. (a) Optical image and (b) SEM micrograph of BZS thin film on quartz substrate, (c) uniform distribution of the constituent elements in the overlay image, and elemental mapping of individual atoms (d) Ba, (e) Zr, (f) S in BZS film.

In the present study, we investigated the FE phases in CVD grown AF-BZS thin film using temperature dependent Raman study and first-principles calculations. We have investigated the ferroic nature and the origin of AF to FE transition by analysing Raman spectra, calculating the macroscopic polarization from first principles calculations for pure BZS and BZS having oxygen (*O*) impurities (*O*-BZS). We have experimentally identified 23 Raman active modes among the possible 57 optical modes along with second order Raman modes and an impurity related phonon. The outcome highlights on the feasibility of achieving AF to local FE phases in thin film of BZS, which can show superior optoelectronic properties as advanced photovoltaic materials.

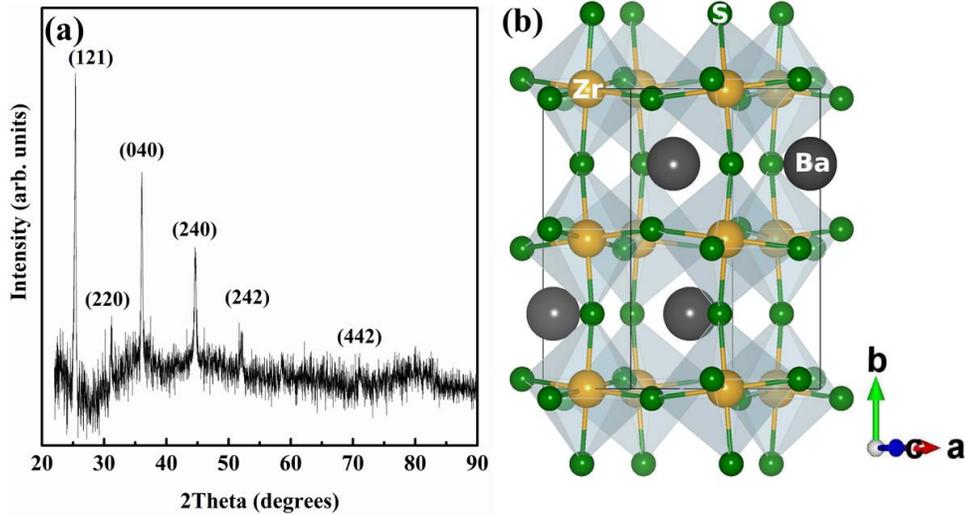

**Figure 2.** (a) XRD pattern of CVD grown BZS thin film, (b) distorted orthorhombic unit cell of BZS perovskite with *Pnma* ($D_{2h}^{16}$) space group.

Optical image and SEM micrograph of the grown thin film is shown in Figure 1a and b, respectively. Here, surface morphology in Figure 1b depicts the polycrystalline nature of the film. Overlay image in Figure 1c and elemental mapping for Ba, Zr and S atom (Figures 1d, e and f, respectively) confirms the uniform stochiometric distribution of atoms in the film. The estimated ratio of Ba:Zr:S atoms is found to be 1:1.2:3.3, which is consistent with the chemical formula of BZS. Phase purity and crystallinity of BZS film are determined by X-ray diffraction (XRD) pattern (Figure 2a) and the peak positions are matched with JCPDS card #00-015-0327. Estimated lattice parameters are *a = 6.984 Å, b = 9.956 Å* and *c = 6.934 Å,* which are slightly smaller than the bulk lattice parameters *a = 7.061 Å, b = 9.977 Å* and *c = 7.014 Å,* possibly due to substrate driven strain in the film.[6] The analysis of XRD pattern shows that the film is grown with the distorted perovskite orthorhombic crystal structure (space group *Pnma* ( $D_{2h}^{16}$ )), shown in Figure 2b. The Wyckoff positions of various atoms are: Ba at $4(c): x, \frac{1}{4}, z$; Zr at $4(b): 0,0, \frac{1}{2}$; and the first sulfur atom $S^I$ at $4(c)$ and second sulfur atom $S^{II}$ at $8(d): x, y, z$. A careful examination of the crystal structure suggest that the unit cell for BZS perovskite contains a network of distorted corner sharing $ZrS_6$ octahedra with Zr atoms placed at the center of each octahedron. Here, Ba atoms are caged centrally between the networks of neighboring $ZrS_6$ units. The octahedral rotations and expansions along with anti-polar displacement of Ba-site from ideal perovskite structure (tolerance factor $t = 1$, the details of $t$ is given in supporting information) often suppress ferroelectricity in perovskites with *Pnma* structure.[14]

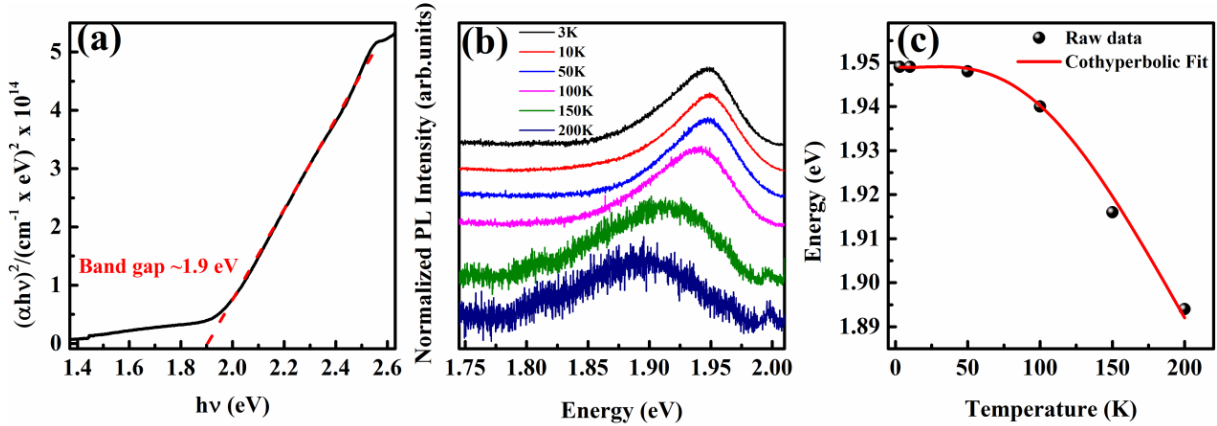

**Figure 3**. Optical characterization of BZS film using (a) absorbance spectra at room temperature with the extrapolated dashed red line showing a direct band gap (~ 1.9 eV), (b) temperature dependent PL spectra, and (c) Energy versus temperature graph, the red line presents hyperbolic cotangent fitting of redshift of PL peak position, which estimates the band gap ~ 1.95 eV at 0 K.

Optical properties of the BZS film have been characterized using absorbance and temperature dependent PL measurements. Figure 3a shows the absorption spectrum of the film, where $\alpha$ is the absorption coefficient and $h\nu$ represents the energy of incident light. Linear dependence of $(\alpha h\nu)^2$ vs $h\nu$ plot near the band edge confirms that the BZS thin film is a direct band gap material with band gap ~ 1.9 eV, at room temperature. Figure 3b shows normalized PL spectrum from a single broad asymmetric emission peak, band gap varying from ~ 1.95 eV to ~ 1.85 eV with increase in temperature. The variation of optical band gap with temperature (Figure 3c), is fitted with standard hyperbolic cotangent, $E(T) = E(0) - C \times E_{ph}\left[\coth\left(\frac{E_{ph}}{k_B T}\right) - 1\right]$, where $E(0)$ is optical band gap at 0 K, $C$ is a dimensionless coupling constant between electron and phonon, $k_B$ is Boltzmann's constant and $E_{ph}$ is the average phonon energy coupling with electron.[15] Here, redshift of the optical band gap is related to electron-phonon coupling and the estimated $E(0)$ is found to be ~ 1.95 eV and $E_{ph}$ = 14.8 meV (~ 120 cm$^{-1}$) for the BZS thin films.

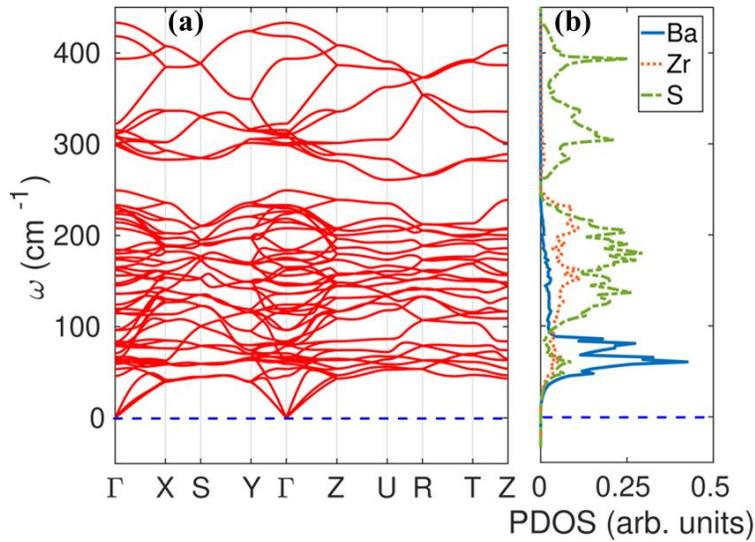

**Figure 4.** (a) Phonon dispersion of BZS film and (b) atom-projected phonon DOS for Ba, Zr and S atoms.

To understand the effect of strain on the vibrational properties of BZS film, we have generated phonon band structure and phonon density of states (PDOS) by using its experimental lattice parameters as shown in Figure 4 a and b respectively. The *Pnma* ($D_{2h}^{16}$) structure of BZS contains four formula units per unit cell resulting in 57 zone-center optical phonons classified as $(7A_g + 5B_{1g} + 7B_{2g} + 5B_{3g}) + (9B_{1u} + 7B_{2u} + 9B_{3u}) + 8A_u$, where only first 24 are Raman active. The next set of 25 modes are infrared active whereas the 8 $A_u$ modes are optically inactive.[16] The calculated Raman active modes are tabulated in Table S1, in supporting information. These modes show close agreement with earlier theoretical work by Gross *et. al.*[17] The dynamical stability of the compound is manifested through the absence of imaginary modes in the phonon band structure (Figure 4a). Atom-projected PDOS reveals that the low-frequency modes (< 100 cm$^{-1}$) have main contributions coming from the vibration of the Ba atoms while the high-frequency modes (> 250 cm$^{-1}$) are dominated by the atomic vibration of the S atoms. However, the mid-frequency phonons involve vibration of both Zr and S atoms.

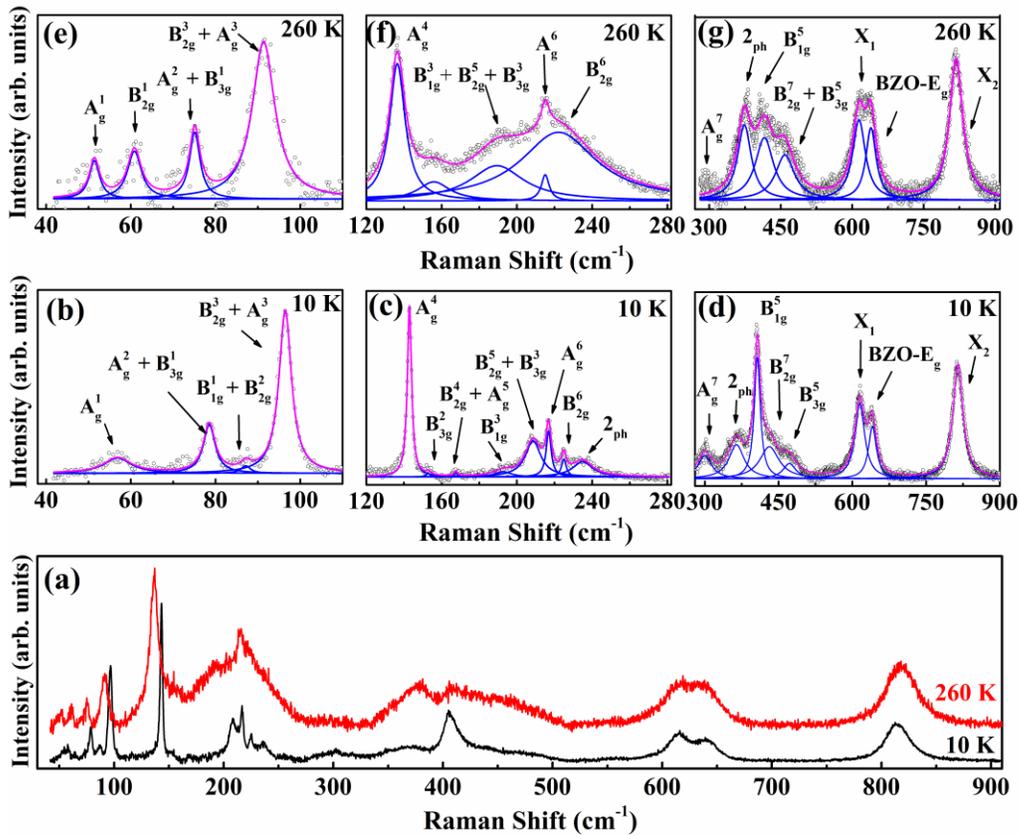

**Figure 5.** Raman spectrum of BZS thin film showing, (a) comparison of spectra at 10 K and 260 K, and Lorentzian fitting of the Raman spectra for the spectral range (b, e) 40 – 120 cm$^{-1}$, (c, f) 120 – 280 cm$^{-1}$ and (d, g) 300 – 900 cm$^{-1}$ at 10 K and 260 K, respectively. Open circles represent raw data, blue solid lines correspond to best-fit results of the individual peak, and pink curve shows the cumulative spectrum.

The Raman spectra at 10 K is quite distinct from the spectra at 260 K (Figure 5a), due to reduced spectral width at low temperature and appearance of new modes. Since the 300 K Raman spectra are little noisy (Figure S2, supporting information), we have chosen the 260 K data. Figures 5 b, c, d show the detailed spectra of 10 K, while Figures 5 e, f, g show the spectra for 260 K in three spectral ranges: (i) 40 to 120 cm$^{-1}$ (ii) 120 to 280 cm$^{-1}$ and (iii) 300 to 900 cm$^{-1}$, respectively. Each spectral range is fitted with multiple Lorentzian peak function to estimate peak position and the origin of peaks has been assigned (tabulated in supporting informations Table S1), according to our phonon calculations shown in Figure 4.

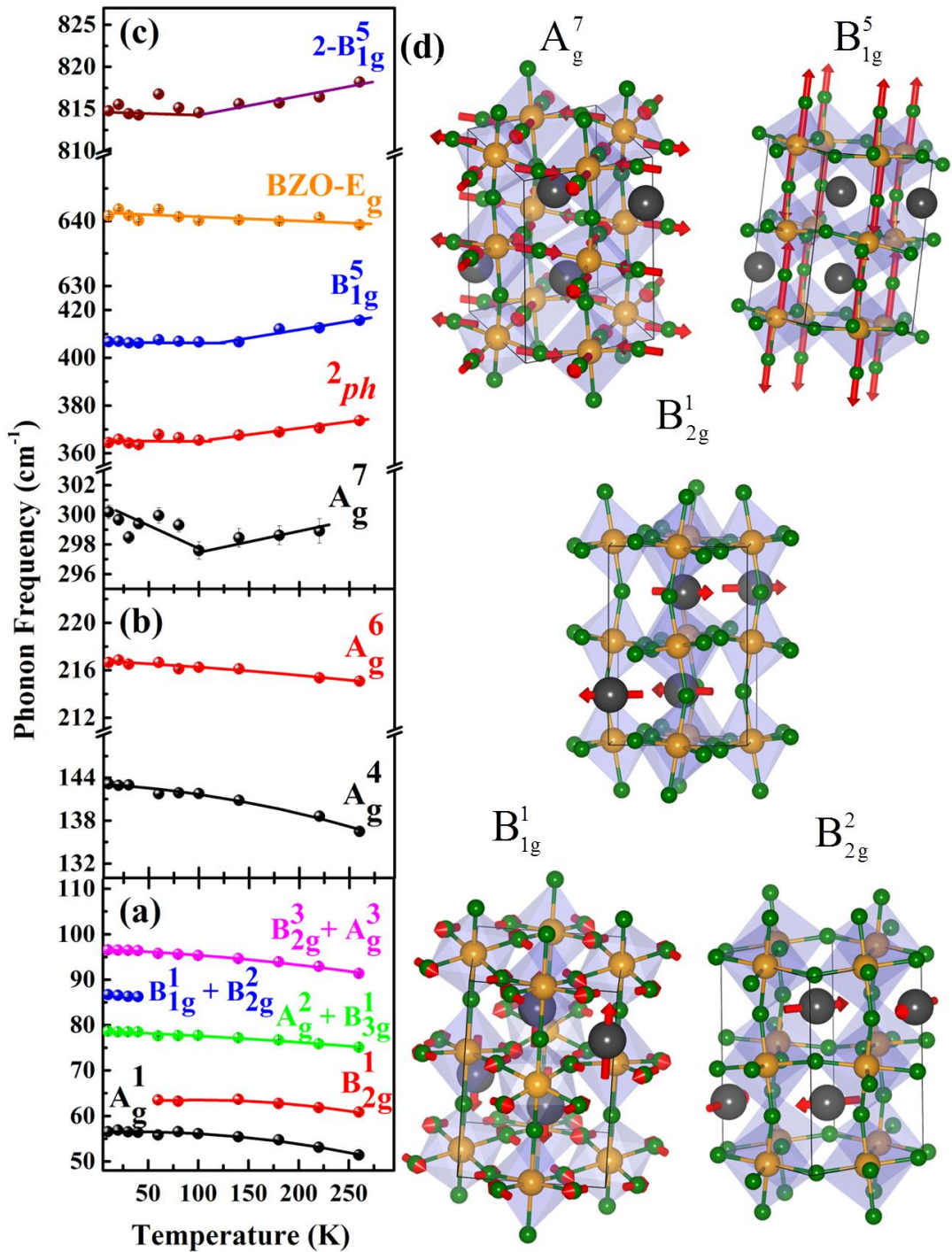

**Figure 6.** Temperature dependence of phonon frequency for the spectral range (a) 40 to 120 cm$^{-1}$ (b) 120 to 280 cm$^{-1}$ (c) 280 to 830 cm$^{-1}$, (d) schematic of the five Raman active modes having anomalous response with temperature.

Temperature dependence of phonon frequency is presented in Figure 6, while the complete spectrum is shown in Figure S1, supporting information. All the Raman modes in spectral range 40-280 cm$^{-1}$ (shown in Figure 6 a and b) are red-shifted due to thermal expansion and corresponding increase in force constant as temperature rises. The reduction in phonon frequency with increasing temperature owes to two major reasons, (i) volume change due to thermal-expansion, and (ii) decay of harmonic mode because of multiphonon interactions.[18] To estimate the thermal renormalization of the phonon frequency, the obtained frequency is fitted by the equation

$$\omega(T) = \omega_0 - A\left[1 + \frac{2}{(e^x - 1)}\right] - B\left[1 + \frac{3}{(e^y - 1)} + \frac{3}{(e^y - 1)^2}\right], \text{ where } x = \frac{\hbar\omega_0}{2k_BT} \text{ and } y = \frac{\hbar\omega_0}{3k_BT},$$

$\omega_0$ is the harmonic frequency at $T = 0$ K, $A$ and $B$ are anharmonic fitting parameters related to probability of three- and four-phonon decay processes.[18] Since the exepression of multi-phonon decay could not be fitted with two-phonon process, we have used the three-and four-phonon terms in the expression. Table S2 in the supporting information gives details of the fitting parameters of the corresponding modes. Interestingly, for bulk BZS, $B_{2g}^1$ (~ 60.8 cm$^{-1}$) is prominently present up to 14 K,[17] however for the thin film discussed here, this mode appears above 60 K where $B_{1g}^1$ and $B_{2g}^2$ (~ 85 cm$^{-1}$) disappear. Generally, such emergence and supression of Raman modes for perovskites signify either modulation in crystal structure or transitions like AF-FE or FE-PE.[19-20] Experimental and theoretical studies show that BZS does not undergoe any structural transition at very low temperatures as well as high pressure.[17] Therefore, such anomalous behavior of Raman active modes cannot be assigned to any structural transition. Here, all the three modes, $B_{1g}^1$, $B_{2g}^2$ and $B_{2g}^1$, involve vibrations of Ba atoms in staggered manner with $B_{1g}^1$ mode has the rotation of octahedra involved as well. Their existence stabilizes the *Pnma* structure against ferroelectricity by local distortion. Schematic of atomic vibration for anomalous modes are shown in Figure 6d. For perovskites belonging to non-polar space group (*Pnma*), the octahedral distortion is considered as the driving mechanism to suppress ferroelectricity by inducing anti-polar displacement of Ba-site cations.[14] If Ba-site distortion is not present then octahedral rotations alone will not be able to stabilize the *Pnma* structure and the system will adopt *R3c* structure, which subsequently may develop FE phases.[14] Since, for perovskites with $t<1$, octahedral rotation of $ZrS_6$ units plays a significant role in AF or FE polarization. Therefore, the cause of transitions can be determined by carefully analysing the nature of Raman modes involved. While most of the modes have shown the usual red-shift of the frequency with increase in temperature, the modes $A_g^7 \sim 300$ cm$^{-1}$, $2_{ph} \sim 373$ cm$^{-1}$, $B_{1g}^5 \sim 415$ cm$^{-1}$ and $X_2 \sim 818$ cm$^{-1}$ have shown anomalous behavior (Figure 6c). Here, the origin of FE transition is established from anomalous behavior of $A_g^7$ mode (~ 300 cm$^{-1}$), which involves in-phase streching of basal S atom, and $B_{1g}^5$ mode (~ 420 cm$^{-1}$), which involves out-of-phase streching of S atom at apical site of $ZrS_6$ octahedra (shown in Figure 6d). The modes at ~ 234.9 cm$^{-1}$ and ~ 365 cm$^{-1}$ are related to two-phonon features (*2ph*) as observed by Gross *et. al.*[12] also. Based on the frequency of the $X_2 \sim 818$ cm$^{-1}$, which is almost twice of $B_{1g}^5 \sim 415$ cm$^{-1}$ and shows identitical anomalous behavior, we assign this mode to the second order of $B_{1g}^5$ mode ($2\text{-}B_{1g}^5$). In addition to these obesrvations, the presence of *O* impurities due to incomplete sulfurization of the precursor $BaZrO_3$ can result in the localised dipole moment by breaking the center of inversion symmetry of the $ZrS_6$ octahedra.[21] Since mode at ~ 641 cm$^{-1}$ has overlap with the frequency of $E_g$ mode of $BaZrO_3$, this mode has been assigned as the impurity related mode in BZS.[20] From the anomalous behavior of $A_g^7$ and $B_{1g}^5$, one may infer the presence of *O* impurity in the BZS film coupled with AF-FE transitions.

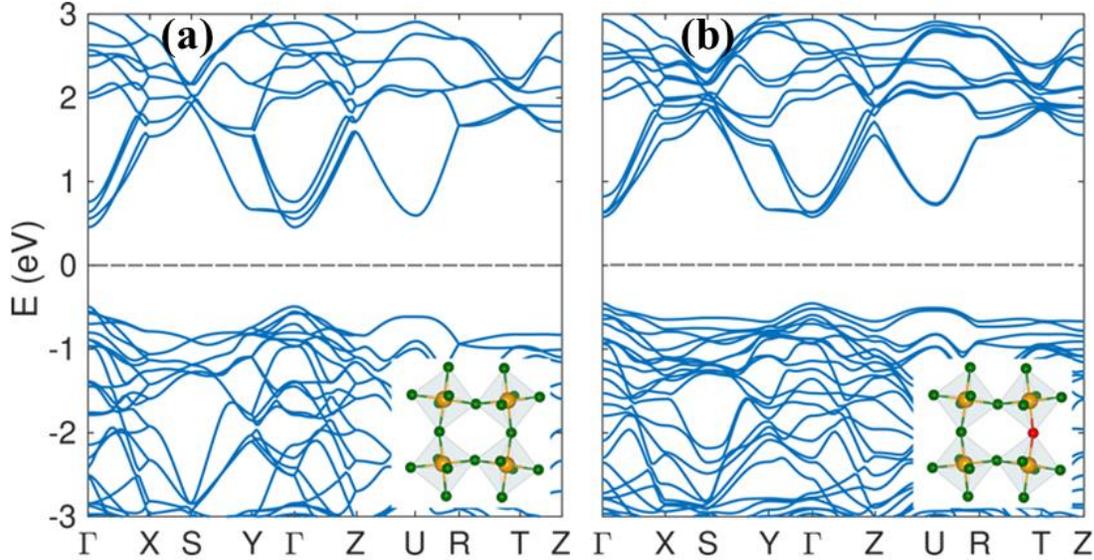

**Figure 7.** Electronic band structure of, (a) thin film BZS, and (b) O-BZS with a single *O* impurity (shown in red color) placed at the apical position of *S* atom (green color) in ZrS$_6$ octahedra.

To understand the role of *O* impurities in BZS thin film, we have calculated electronic band structure of pure BZS and *O*-BZS (*O* atoms displayed with red sphere), as shown in Figure 7. The crystal structures of pure BZS and *O*-BZS with *O* atom at single apical site of S atom are shown in the inset Figure 7 a and b, respectively. The electronic band structure of pure BZS (Figure 7a) shows that the material is a semiconductor with direct band gap ~ *0.95 eV*. Since, the DFT calculations underestimate the band gap thus the band gap value obtained within GGA is reasonable with the experimental observations. Further, even with the presence of *O* impurities, the material remains a direct band gap semiconductor with band gap value ~ 1.03 eV as shown in Figure 7b. Interestingly, our BZS film has a band gap ~ 1.9 eV which is slightly higher than the band gap of bulk BZS ~ 1.85 eV,[6] possibly due to the localised *O* impurities. Our experimental and theoretical studies show that localised *O* impurities do not modulate the electronic properties of BZS dramatically. We have also investigated the role of any S vacancy that can develop ferroelectricity in BZS. Interestingly, our calculations suggest that the presence of S vacancy modulates electronic band structure and develops metallicity in BZS (shown in Figure S3, supporting information). Moreover, since pure BZS is AF due to the presence of a center of inversion, we have calculated macroscopic polarization to investigate the ferroic nature of pure-BZS and *O*-BZS. Our calculations suggest that pristine-BZS is AF as |P| = 0, whereas, *O*-BZS loses the center of inversion symmerty and develops a net FE polarization with |P| = 0.26 µc/cm$^2$. Several TMCPs such as $Ca_3Zr_2S_7$ and $Ca_3Hf_2S_7$ have high FE polarization, however, the high band gaps of such compounds make them unsuitable for photovoltaic application.[23] On the contrary, chalcogenide perovskite having FE domains along with optimum optical band gap provide an additional route for reduced charge recombination and improved charge transport leading to an efficient photovoltaic applicaiton. Thus, our experimental findings supported by theoretical calculations reveal the possible mechanism in obtaining nanoscale FE response in AF-BZS thin film.

**Conclusions**

In summary, we have shown localised FE phases arising from the loss of center of inversion symmetry in AF BZS thin film at nanoscale. The local FE is implied by the anomalous softening of high frequency Raman active modes $A_g^7$ ~ 300 cm$^{-1}$ and $B_{1g}^5$ ~ 420 cm$^{-1}$ involving the vibration of S atoms of the $ZrS_6$ octahedra. The absence of imaginary modes in our phonon band structure also confirms that the BZS thin film has dynamically stable *Pnma* structure. Further, *O* impurities give local FE but do not modulate the nature and the magnitude of electronic band gap. For pristine BZS, there is no macroscopic electronic polarization (|P| = 0), whereas BZS with *O* impurity has |P| = 0.26 µc/cm$^2$, implying the FE nature of the thin film, thus making it a promising FE photovoltaic material.

**Experimental**

Thin film of BZS was grown over quartz substrate by CVD technique. In the process, pre-deposited $BaZrO_3$ film over quartz substrate was treated with a mixture of $N_2$ and $CS_2$ vapors for sulfurization, which resulted in continuous BZS film. Details of the CVD process and the growth parameters were described in earlier work.[7] Optical image of the BZS thin film was obtained using Olympus BX53 microscope with 100X objective and the morphology was characterized by JFEI, Nova Nano SEM-450, field emission scanning electron microscopy (FESEM) at 15 kV acceleration voltage. Elemental composition of the film was mapped by Energy Dispersive Spectroscopy (EDS) coupled with FESEM. To assess the crystal structure and phase purity, XRD was done using rotating anode Rigaku Smartlab diffractometer in Bragg-Brentano geometry with CuKα radiation ($\lambda = 1.5406$ Å). Micro-Raman measurements were performed using Horiba Jobin-Yvon LabRAM HR evolution Raman spectrometer in back scattering geometry with He-Ne laser excitation ~ 633 nm (for near resonant Raman) and solid state laser excitation ~ 785 nm (for non-resonant Raman), 1800 gr/mm grating and Peltier cooled CCD detector. Micro-PL study was performed using the same Raman spectrometer with 532 nm (2.3 eV) laser excitation and 600 gr/mm grating. Temperature dependent Raman and PL were performed using Montana cryostat in the range of 3 - 300 K. Optical band gap of the film was estimated using Perkin Elmer UV/Vis/NIR spectrometer Lambda 750 at room temperature in absorbance mode.

First-principle calculations, based on density functional theory (DFT),[24-25] were performed by using a plane-wave basis set and projector-augmented wave (PAW) potentials,[26-27] as implemented in the Vienna *ab-initio* simulation package (VASP).[28] The wave functions were expanded in the plane-wave basis with a kinetic energy cutoff of 500 eV. For the exchange-correlation function, we used generalized gradient approximation (GGA) with the Perdew-Burke-Ernzerhof (PBE).[29] The reciprocal space integration was carried out with a Γ-centered k-mesh of 8×6×8. During structural relaxations, positions of the ions were relaxed until the Hellman-Feynman forces became less than $10^{-3}$ eV/Å. Phonons were calculated from density functional perturbation theory (DFPT) as implemented in the PHONOPY code.[30] The macroscopic electronic polarization has been calculated by using the Berry phase expressions.[31]

- **ASSOCIATED CONTENT**

**Supporting Information**

The supporting information has discussion on the tolerance factor (*t*), temperature dependent Raman spectra in Figure S1, list of experimentally observed Raman modes and theoretically estimated phonon frequencies in Table S1, the details of fitting parameter estimated by three and four phonon decay process for Raman active modes in Table S2. Figure S2 has resonant and non-resonant Raman spectra, at 300 K. Figure S3 has electronic band structure of BZS with S vacancy.

- **AUTHOR INFORMATION**

**Corresponding Author**

Ajay Soni: School of Basic Sciences, Indian Institute of Technology Mandi, Mandi-175075, Himachal Pradesh, India, ORCID: 0000-0002-8926-0225, *Email: ajay@iitmandi.ac.in

**ORCID:**

Juhi Pandey: 0000-0003-4515-3062

Debjit Ghoshal: 0000-0003-3204-0755

Dibyendu Dey: 0000-0002-2639-3266

Tushar Gupta: 0000-0002-2161-2925

Arghya Taraphder: 0000-0002-2916-1532

Nikhil Koratkar: 0000-0002-4080-3786

**Notes**

The authors declare no competing financial interest.

## ACKNOWLEDGEMENTS

A.S. acknowledges DST-SERB India (Grant No. CRG/2018/002197) for funding support and IIT Mandi for research facilities. NK acknowledges funding support from the USA National Science Foundation (Award Number 2013640).

# Supporting Information

# Ferroelectric Polarization in Antiferroelectric Chalcogenide Perovskite BaZrS$_3$ Thin Film

Juhi Pandey,[1] Debjit Ghoshal,[2] Dibyendu Dey,[3,4] Tushar Gupta,[5] Arghya Taraphder,[3] Nikhil Koratkar,[5] and Ajay Soni[1]*

[1]School of Basic Sciences, Indian Institute of Technology Mandi, Mandi-175075, Himachal Pradesh, India

[2]Department of Chemical and Biological Engineering, Rensselaer Polytechnic Institute, Troy-12180, New York, USA

[3]Department of Physics and Centre for Theoretical Studies, Indian Institute of Technology, Kharagpur-721302, West Bengal, India

[4]Department of Physics, Arizona State University, Tempe-85287, Arizona, USA

[5]Department of Mechanical, Aerospace and Nuclear Engineering, Rensselaer Polytechnic Institute, Troy-12180, New York, USA.

Corresponding Author Email: ajay@iitmandi.ac.in

The supporting information has description about the tolerance factor, temeprature dependent Raman spectra, assignments of experimentally observed Raman active modes from phonon calclulations, Resonant and non-resonant Raman experiments, list of fitting parameters estimated from three and four phonon decay process for variation of phonon frequency with temperature and electronic band structure for BZS film having *S* vacancy.

1. **Tolerance factor**

In general, the degree of distortion from an ideal perovskite structure ($ABX_3$) can be estimated from empirical Goldschmidt *tolerance factor,*[1] *t*, which depends on the ionic radii of *A, B* and *X* atoms as described by $t = \frac{R_{A-X}}{R_{B-X}\sqrt{2}}$. Here, $R_{A-X}$ is sum of the ionic radii of *A* and *X* atoms and $R_{B-X}$ is sum of *B* and *X* ionic radii.[2] If $t > 1$ due to large ionic size of A cation compared to B cation, the *B-$X_6$* octahedra are stretched from their *B-X* bond length, leading to off-centering of B cation and a certain dipole moment developed. Therefore, the perovskites with $t > 1$ are usually FE in nature. On the other hand, for *t < 1,* because of smaller ionic size of A cation, $BX_6$ octahedra tilts to fill the space to suppress ferroelectricity. Thus, octahedral rotations and expansions owing to *t < 1* often generate a low temperature AF phase. The ferroelectricity is suppressed in most of the perovskites with *Pnma* space group because of displacement of A-site cation from ideal perovskite structure (*t = 1*) due to rotation of *$BX_6$* octahedra.[3] DFT calculations performed by Bennet *et. al.*[4] determines the ground state atomic structure of BZS as AF with *t = 0.95*.

2. **Temperature dependent Raman spectra**

Figure S1 shows the temperature dependent Raman spectra for the three spectral ranges (a) 40 – 120 cm$^{-1}$, (b) 120 – 280 cm$^{-1}$ and (c) 280 – 900 cm$^{-1}$ from 3 K to 300 K. The $B_{2g}^1 \sim 60$ cm$^{-1}$ disappears below 60 K and $B_{1g}^1 + B_{2g}^2 \sim 80$ cm$^{-1}$ appears below 60 K. $B_{2g}^1$ and $B_{1g}^1 + B_{2g}^2$ are shown in black and red dashed rectangles respectively.

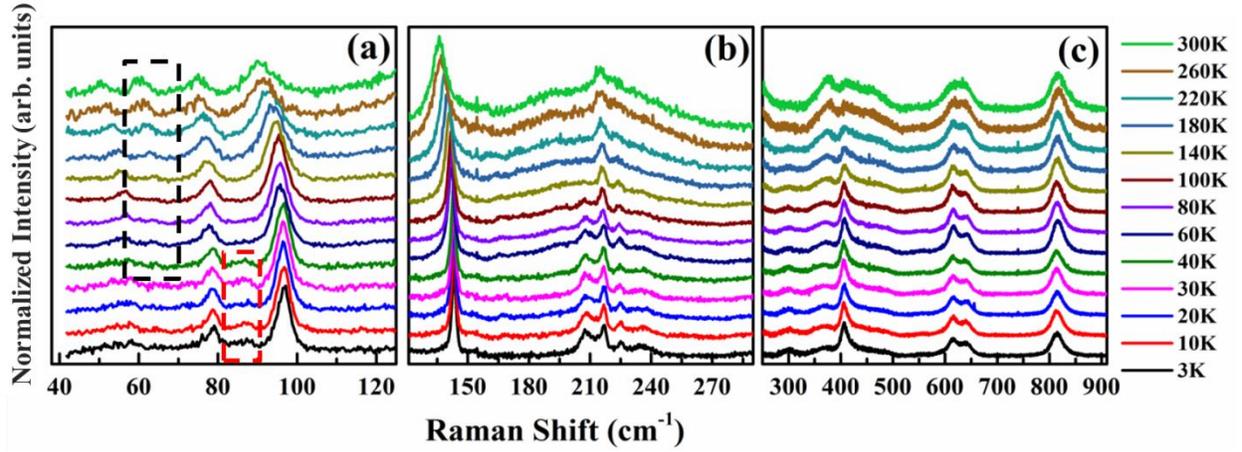

**Figure S1.** Temperature dependent Raman spectra from 3 K to 300 K for spectral ranges (a) 40 – 120 cm$^{-1}$, (b) 120 – 280 cm$^{-1}$ and (c) 280 – 900 cm$^{-1}$.

## 3. Assignment of symmetry of experimentally observed Raman active modes

Table S1 has the theoretical frequency ($\omega_{theory}$) calculated using DFT for BZS film and experimental frequency ($\omega_{exp}$) observed through Raman spectroscopy at 10 K. The symmetry of the Raman active modes is determined through DFT calculation.

**Table S1.** Raman vibrational mode assignment based on our DFT calculation and experimental observation of phonon frequencies at 10 K

| Mode | $\omega_{theory}$ (cm$^{-1}$) | $\omega_{exp}$ (cm$^{-1}$) | Mode | $\omega_{theory}$ (cm$^{-1}$) | $\omega_{exp}$ (cm$^{-1}$) | Mode | $\omega_{theory}$ (cm$^{-1}$) | $\omega_{exp}$ (cm$^{-1}$) |
|---|---|---|---|---|---|---|---|---|
| $A_g^1$ | 59.3 | 56.52 | $B_{1g}^2$ | 116.2 | - | $A_g^6$ | 218.4 | 217.2 |
| $B_{2g}^1$ | 63.8 | - | $A_g^4$ | 147.6 | 143.1 | $B_{2g}^6$ | 226.4 | 225.0 |
| $A_g^2$ | 78.7 | 78.6 | $B_{3g}^2$ | 164.6 | 154 | $B_{3g}^4$ | 299 | 285 |
| $B_{3g}^1$ | 80.9 | 78.6 | $A_g^5$ | 173.6 | 168.3 | $B_{1g}^4$ | 309.2 | 285 |
| $B_{1g}^1$ | 82.4 | 86.7 | $B_{2g}^4$ | 175.7 | 168.3 | $A_g^7$ | 322.5 | 299.7 |
| $B_{2g}^2$ | 83.9 | 86.7 | $B_{1g}^3$ | 183.4 | 192 | $B_{1g}^5$ | 393.6 | 407 |
| $A_g^3$ | 96.2 | 96.5 | $B_{2g}^5$ | 207.8 | 208.5 | $B_{2g}^7$ | 418.3 | 428 |
| $B_{2g}^3$ | 96.5 | 96.5 | $B_{3g}^3$ | 214.5 | 208.5 | $B_{3g}^5$ | 433 | 428 |

## 4. Resonant and non-resonant Raman studies to determine the origin of unidentified modes

Figure S2 shows the near resonant and non-resonant Raman spectra upon excitation with 633 nm and 785 nm respectively at room temperature. Generally, the relative intensity of first order modes is significantly large compared to second and higher order modes. However, in our case, the relative intensity of first and second order modes is comparable probably due to near-resonant Raman scattering with 633 nm (1.95 eV) excitation. In the case of resonant Raman scattering, where the energy of excitation laser is comparable with the optical band gap, intense second as well as higher order modes can be observed.[5-6] On the other hand, with non-resonant excitation, all the processes higher than second order disappear. The intensity of second order modes are suppressed significantly on excitation with 785 nm laser. Raman spectrum of quartz is obtained using 633 nm excitation. The origin of several high order Raman modes observed above 500 cm$^{-1}$ is determined based on the investigation of near resonant and non-resonant Raman spectra at room temperature.

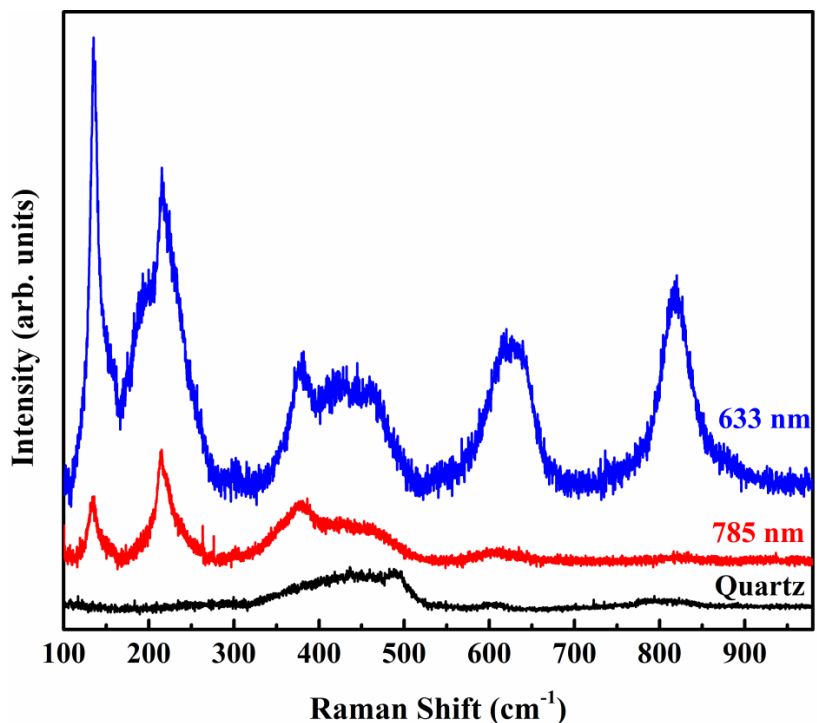

**Figure S2.** Raman spectra at 300 K on excitation with 633 nm for BZS film (blue color), quartz (black color) and 785 nm excitation of BZS film (red color.)

## 5. Table S2 tabulate the fitting parameter estimated by fitting with the three and four phonon decay process

Table S2. List of fitting parameter determined for temperature dependent variation of phonon frequency.

| Mode | $\omega_0$ (cm$^{-1}$) | A (cm$^{-1}$) | B (cm$^{-1}$) |
|---|---|---|---|
| $A_g^1$ | 56.4 | -0.0222 | 0.0006 |
| $B_{2g}^1$ | 62.4 | -0.07885 | 0.0008418 |
| $A_g^2 + B_{3g}^1$ | 78.68 | 0.04037 | 0.0002185 |
| $B_{1g}^1 + B_{2g}^2$ | 96.62 | 0.04439 | 0.0008312 |
| $A_g^4$ | 143.1 | 0.06795 | 0.002527 |
| $A_g^6$ | 216.8 | 0.05785 | 0.0007162 |

## 6. Electronic band structure of BaZrS$_3$ with S vacancy

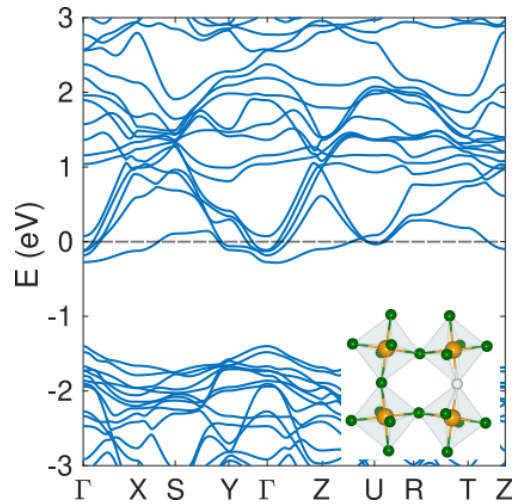

**Figure S3.** Electronic band structure of BaZrS$_3$ having S vacancy showing metallic nature.

Figure S3 represents the calculated electronic band structure for BaZrS$_3$ having S vacancy. In the presence of S vacancy, the system becomes metallic due to shift in chemical potential. Metallic

BZS does not show any polarization thus may not be suitable for photovoltaic and optoelectronic applications.

**References:**

...